# Predicting interstitial elements in Refractory Complex Concentrated Alloys

Aomin Huang[1], Siya Zhu[1], Calvin Belcher[2], Ryker Rigsby[1], Diran Apelian[2], Raymundo Arróyave[1,3,#], Enrique J. Lavernia[1,3,#]

[1]Department of Materials Science and Engineering, Texas A&M University, College Station, TX-77843, USA
[2]Department of Materials Science and Engineering, University of California, Irvine, Irvine, CA 92617, USA
[3]J. Mike Walker '66 Department of Mechanical Engineering, Texas A&M University, TX-77845, USA

Corresponding authors #:
rarroyave@tamu.edu
lavernia@tamu.edu

**Abstract**
Refractory complex concentrated alloys (RCCAs), composed of multiple principal refractory elements, are promising candidates for high-temperature structural applications due to their exceptional thermal stability and high melting points. However, their mechanical performance is often compromised by interstitial impurities—particularly oxygen, nitrogen, and carbon—which segregate to grain boundaries and promote embrittlement. In this study, we investigate the solubility and thermodynamic behavior of oxygen interstitials in a model NbTiHfTa RCCA system. We synthesized NbTiHfTa alloys with varying oxygen contents (0–5 at.%) via plasma arc melting and characterized their phase evolution and microstructure using XRD, SEM, and TEM. Complementary computational modeling was performed using machine-learning interatomic potentials (MLIPs) integrated with Monte Carlo simulations to probe oxygen interactions at the atomic scale. Our results reveal a solubility limit for oxygen between 0.8 and 1.0 at.%, beyond which $HfO_2$ formation is energetically favorable. This combined experimental–computational framework provides a predictive approach for managing interstitial behavior in RCCAs, enabling improved alloy design strategies for enhanced mechanical performance.

**Keywords**
Refractory complex concentrated alloys
Interstitials

# 1. Introduction

   Complex concentrated alloys (CCAs) represent a novel alloy design strategy that involves mixing three or more metallic elements in near-equiatomic ratios. [1,2]. This alloy design significantly expands the compositional space for tailoring material properties by leveraging factors such as lattice distortion and configurational entropy. Early studies focused primarily on CCAs composed of 3d transition metals, which typically formed single-phase face-centered cubic (FCC) structures known for their excellent damage tolerance and fracture resistance. However, these alloys exhibit rapid performance degradation at elevated temperatures due to thermal softening, limiting their suitability for high-temperature applications. To meet the growing demand for structural materials capable of operating in extreme thermal environments, a subclass of CCAs composed predominantly of refractory elements—refractory complex concentrated alloys (RCCAs) [3]—has been developed. These alloys leverage the high melting points of their constituent elements to maintain mechanical strength at elevated temperatures. However, the significant differences in melting points among these elements can lead to the formation of inhomogeneous microstructures, such as dendritic and interdendritic regions, during solidification. Such heterogeneities may promote stress localization at interfaces, increasing the risk of premature mechanical failure.[4]. Cold rolling followed by heat treatment is a commonly employed strategy to reduce microstructural inhomogeneities; however, this process requires adequate room-temperature deformability—a property that most RCCAs inherently lack. This limitation is further exacerbated by the strong affinity of RCCAs for interstitial impurities such as oxygen, nitrogen, and carbon. Due to their relatively small atomic radii, these elements readily occupy interstitial sites within the metal lattice and tend to segregate at grain boundaries, acting as primary contributors to grain boundary embrittlement and diminished workability.

   A review of the scientific literature indicates that research on RCCAs has predominantly focused on two model systems: MoNbTaW and HfNbTaTiZr [3,5–12], with significant differences found in their room-temperature ductility. These differences have been attributed to the effects of the constituent elements on dislocation motion and their sensitivity to interstitial impurities such as oxygen, nitrogen, and carbon. The NbTaTiHfZr alloy and similar RCCAs, which were inspired by the HfNbTaTiZr system, were found to have relatively low strength but excellent ductility at room temperature [13–15]. A follow-up study on the $Nb_{45}Ta_{25}Ti_{15}Hf_{15}$ RCCA was developed by Zhang et al. [14] to achieve moderate strength and high room-temperature ductility, with low sensitivity to interstitial impurities. The as-cast $Nb_{45}Ta_{25}Ti_{15}Hf_{15}$ alloy can undergo a thickness reduction of over 90% during cold rolling. The composition of the NbTaTiHf alloy, as well as other alloys in the NbTaTiHfZr RCCA system, is similar in composition and structure to the industrially developed C-103 alloys, which are comprised of Nb–10 wt. % Hf and 1 wt. % Ti and has a low sensitivity to O [16]. In the NbTaTiHf alloy, oxygen sensitivity has not yet been quantified or fully understood. Therefore, the goal of this work is to measure the NbTaTiHf RCCA's sensitivity to oxygen and investigate how oxygen affects the alloy's microstructure. The presence of impurities in RCCAs is influenced by their quantities, solubility, and the interaction energy with base metals [17]. Due to the complexity of the local chemical environment, predicting impurities and their solubility in RCCAs remains challenging, posing a significant challenge for alloy design when interstitial elements are involved. Although oxygen is widely believed to be detrimental to the mechanical performance of RCCAs, published studies, for example, by Lu and his group, have reported contradictory findings of enhanced strength and ductility via intentionally doping oxygen into RCCAs [18–21]. Accordingly, the development of accurate predictive models is critical for alloy design, as they provide fundamental insight into the influence of impurities on the atomic structure and mechanical properties of RCCAs.

In this study, Nb$_{45}$Ta$_{25}$Ti$_{15}$Hf$_{15}$ refractory complex concentrated alloys (RCCAs) containing 0–5 at.% oxygen were synthesized using plasma arc melting. High-purity elemental feedstocks, including Nb, Ta, Ti, Hf, and Nb$_2$O$_5$, were used as starting materials, and repeated melting and flipping were performed under high vacuum with argon backfilling to ensure chemical homogeneity and minimize contamination. The resulting ingots were analyzed using inert gas fusion to quantify bulk oxygen content, confirming that the targeted compositions were closely achieved. Phase identification via X-ray diffraction revealed a single BCC phase in the base alloy and the emergence of secondary phases—attributable to HfO$_2$—at oxygen concentrations ≥0.5 at.%. Cold rolling experiments were conducted to assess room-temperature deformability as a function of oxygen content. Alloys containing more than 0.5 at.% O exhibited rapid embrittlement, with surface cracking occurring at thickness reductions of less than 10%, suggesting a loss of ductility due to either interstitial strengthening or oxide formation. Detailed microstructural characterization using SEM, TEM, and EDS confirmed that at low oxygen concentrations (e.g., 0.5 at.%), oxygen is uniformly distributed in solid solution, while higher concentrations (≥1 at.%) promote the formation of grain boundary HfO$_2$ precipitates and chemical segregation.

To complement our experimental findings, we developed a computational framework to predict the solubility limit and configurational behavior of interstitial oxygen in NbTaTiHf RCCAs. Traditional approaches based on density functional theory (DFT) or Special Quasirandom Structures (SQS) are limited by system size and fail to capture local chemical ordering and interstitial interactions. While cluster expansion (CE) methods offer better scalability, they require extensive training data to accurately model complex interstitial environments. To overcome these challenges, we employed universal machine-learning interatomic potentials (MLIPs) within the MaterialsFramework [22], enabling high-throughput evaluation of RCCA+O configurations with near-DFT accuracy. By statistically analyzing hundreds of randomly generated structures and applying a Monte Carlo sampling scheme, we identified favorable oxygen environments and observed a strong preference for coordination with Ti and Hf. These results, in agreement with experimental characterization, establish the oxygen solubility limit in NbTaTiHf RCCAs to be between 0.8 and 1.0 at.%.

## 2. Experimental

Ingots (~30 g each) of Nb$_{45}$Ta$_{25}$Ti$_{15}$Hf$_{15}$ refractory complex concentrated alloys (RCCAs) containing 0, 0.5, 1, 2, and 5 at.% oxygen were synthesized via plasma arc melting using an Arc200 arc melter (Arcast Inc., Oxford, ME, USA). The alloy charges were prepared from elemental Nb, Ta, Ti, and Hf slugs, along with Nb$_2$O$_5$ lump feedstock (all ≥99.95% purity, metal basis), procured from Alfa Aesar (Ward Hill, MA, USA). To ensure compositional uniformity, the feedstocks were arc-melted into an initial ingot, flipped, and re-melted sequentially five times without compromising the Ar atmosphere. The chamber was evacuated to a base pressure of ~1 × 10$^{-5}$ torr and backfilled with ultra-high-purity argon to suppress oxidation during melting. Care was taken to maintain a stable arc and minimize contamination from the melting environment.

The as-cast buttons were subsequently subjected to grinding and polishing to remove sharp edges and ensure uniform thickness and deformation for the following rolling. Here after the samples were rolled at room temperature with a reduction in thickness per pass of approximately 0.5% until cracks were observed. The bulk oxygen content in each as-cast alloy was quantified via inert gas fusion (IGF), using an Eltra Scientific Elementrac ONH-p2 analyzer (Verder Scientific, Philadelphia, PA, USA) operating at up to 6000 W. Samples were polished with 600-grit SiC paper, sectioned into ~0.1000 g pieces using clean bolt cutters, and analyzed using Ni fluxes and helium carrier gas according to ASTM Standard E3346-22. Triplicate measurements were performed for each alloy composition to minimize statistical error. Preliminary microstructural and compositional characterization was conducted on as-cast specimens using a combination of FEI Quanta 600 FE-SEM and FEI-Heilos NanoLab DualBeam. Initial microstructural analysis was performed with electron back scatter diffraction (EBSD) with a step size ranged from 0.3 to 0.5 μm. For high-resolution analysis, thin lamellae were extracted using focused ion-beam. The regions of interest were protected with sequential electron- and ion-beam-deposited Pt layers (~2 μm total thickness) to prevent surface irradiation damage during milling. Lift-out specimens were attached to Cu grids and progressively thinned using stepwise voltage reduction (30, 15, 10, and 5 kV), followed by final surface cleaning in a precision ion polishing system (PIPS) to remove amorphous layers. The cleaned lamellae were examined using a double-corrected Titan Themis³ 300 S/TEM for atomic-scale imaging and elemental analysis.

## 3. Modeling Methodology

Simulating interstitial oxygen in disordered, multicomponent systems such as NbTaTiHf RCCAs using conventional density functional theory (DFT) poses significant challenges due to high computational costs and limitations on accessible system sizes. Prior studies have employed Special Quasirandom Structures (SQS) to approximate chemical disorder in RCCAs [23–25]; however, this approach has critical drawbacks. Most notably, system size constraints result in unrealistically high interstitial concentrations—for example, a 48-atom SQS with one oxygen atom corresponds to ~2 at.% O, well above the experimentally observed solubility limit. Even with larger supercells (e.g., 120 atoms), the effective oxygen content remains too high to meaningfully study dilute solubility. Moreover, the SQS framework assumes fully random atomic arrangements and therefore cannot capture the local chemical environments and short-range ordering effects introduced by oxygen. In practice, interstitial oxygen tends to induce non-random clustering of specific metallic species, particularly Ti and Hf, which are not represented in idealized SQS models. Furthermore, SQS-based approaches offer little insight into the configurational thermodynamics of multiple interstitials—for example, whether oxygen atoms prefer to cluster or remain isolated—limiting their utility in studying impurity–impurity interactions.

Cluster expansion (CE) methods provide a more scalable alternative, allowing rapid energy evaluations of large configurational spaces and enabling Monte Carlo sampling of low-energy structures. However, the accuracy of CE models depends heavily on the quality and diversity of the training dataset. For chemically homogeneous alloys, models incorporating two-body clusters up to the third or fourth nearest neighbors—alongside select three- and four-body terms—can often achieve good accuracy when trained on small cells [26–28]. Interstitial systems, in contrast, are more complex: modeling the energetics of oxygen occupying octahedral sites (coordinated by six metallic atoms) requires significantly larger and more chemically diverse supercells to resolve the relevant interactions and chemical preferences.

To overcome these limitations, we employed universal machine-learning interatomic potentials (MLIPs), implemented within the newly developed MaterialsFramework, to model RCCA+O systems. MLIPs offer a practical compromise between fidelity and scalability, enabling high-throughput structural relaxations and energy evaluations at near-DFT accuracy but with orders-of-magnitude lower computational cost. All DFT reference calculations were performed using the Vienna Ab initio Simulation

Package (VASP) [29,30] with the Perdew–Burke–Ernzerhof (PBE) exchange–correlation functional and PAW pseudopotentials within the generalized gradient approximation (GGA) [31]. A k-point density of 8000 points per reciprocal atom was used, and the plane-wave energy cutoff was set to 1.3× the ENMAX of the species present.

For MLIP-based calculations, we used the pre-trained ORB-v2 universal potential [32] without further fine-tuning. Structural relaxations and free energy calculations were carried out using the FIRE algorithm [32], implemented within the Atomic Simulation Environment (ASE) and integrated into the *MaterialsFramework* package. To statistically evaluate the preferred local chemical environments of oxygen, we generated and analyzed a dataset of 500 randomly sampled 128-atom RCCA supercells, each containing a single interstitial oxygen atom. Comparison with DFT benchmarks confirmed the accuracy of the MLIP approach, providing confidence in its predictive capability across the relevant compositional and configurational space.

To further explore configurational space and capture the thermodynamic behavior of interstitial oxygen beyond static sampling, we developed a custom Monte Carlo algorithm tailored to RCCA systems with interstitial impurities. While the random supercell approach provided insight into local chemical preferences of isolated O atoms, it does not account for cooperative effects, such as impurity clustering or competition among low-energy configurations at finite concentrations. The Monte Carlo framework was therefore designed to efficiently navigate this high-dimensional energy landscape, leveraging the speed and accuracy of machine-learning interatomic potentials to search for energetically favorable atomic arrangements across both metallic and interstitial sublattices.

In this work, we developed a custom Monte Carlo algorithm designed to identify low-energy atomic configurations in RCCA systems containing interstitial impurities. The framework integrates with machine-learning interatomic potentials to enable efficient structural relaxation and energy evaluation. The simulation begins with an initial configuration, which may be a fully randomized arrangement of metallic atoms and interstitials or a partially optimized structure from a previous run. At each Monte Carlo step, one of two operations is randomly selected: (1) exchanging the positions of two metallic atoms of different species, or (2) swapping the occupations of two interstitial sites—either between different interstitial species or between an interstitial and a vacancy. To enhance sampling efficiency, the selection probabilities are weighted at 70% for metallic site swaps and 30% for interstitial swaps, based on empirical optimization; these weights are user-adjustable within the code.

After each attempted swap, structural relaxation and energy calculations are performed using the machine-learning potential. The resulting energy is then compared with that of the structure prior to the swap. If the energy is reduced, the swap is accepted; otherwise, it is accepted with a probability determined by the Metropolis criterion:

$$p = exp(\frac{E^0 - E}{k}) \qquad (1)$$

where $E^0$ is the energy before swap, $E$ is the energy after swap, and $k$ is a constant specified by users.

In addition, we set a penalty for unaccepted interstitial swaps: the probability of swapping interstitials decreases by 10% for each unaccepted swap in a row and is reset to 30% after an accepted swap. The MC process can be stopped manually, or it stops when the target number of steps (accepted swaps) is reached or the time limit is met. The entire workflow is shown in Fig. 1.

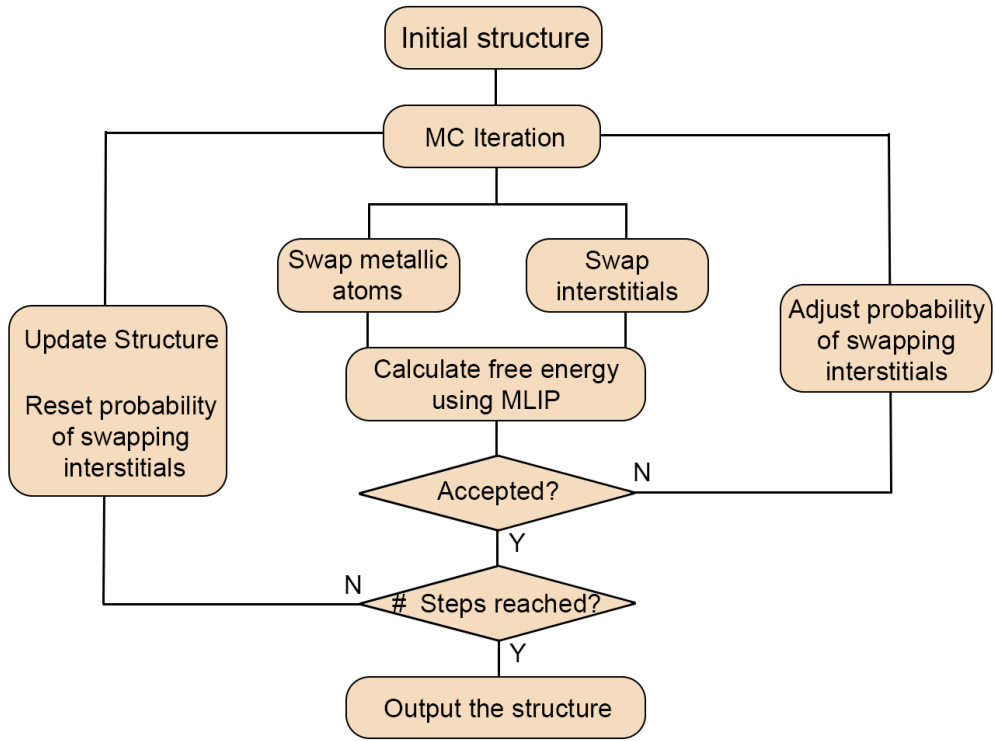

Fig. 1. Flowchart of the Monte Carlo method searching for the ground state structure of RCCA+O

## 4. Results and Discussions

### 4.1 Experiments

The NbTaTiHf alloys were synthesized, targeting 0, 0.5, 1, 2, and 5 at. % O, while maintaining the original atomic fractions of $Nb_{45}Ta_{25}Ti_{15}Hf_{15}$ in the metal constituent elements. The targeted and actual contents of O in the NbTaTiHf alloys, as measured via inert gas fusion, are presented in Table. 1 below. The O content in the as-cast NbTaTiHf RCCAs is on target, with some deviation up to 0.1 at. % O. The alloy with no intentionally added O contained approximately 400 ppm O (0.3 at. % O). For phase composition analysis, the XRD patterns of the as-cast NbTaTiHf alloys are shown below in Fig. 2.

**Table 1.** Targeted and actual contents of O in the experimentally synthesized NbTaTiHf alloys.

| Target O [at. %] | Actual [ppm] | Actual [at. %] |
|---|---|---|
| 0 | 404 ± 37 | 0.30 ± 0.03 |
| 0.5 | 717 ± 120 | 0.54 ± 0.07 |
| 1 | 1388 ± 234 | 1.04 ± 0.04 |
| 2 | 2817 ± 82 | 2.09 ± 0.01 |
| 5 | 6817 ± 202 | 4.90 ± 0.18 |

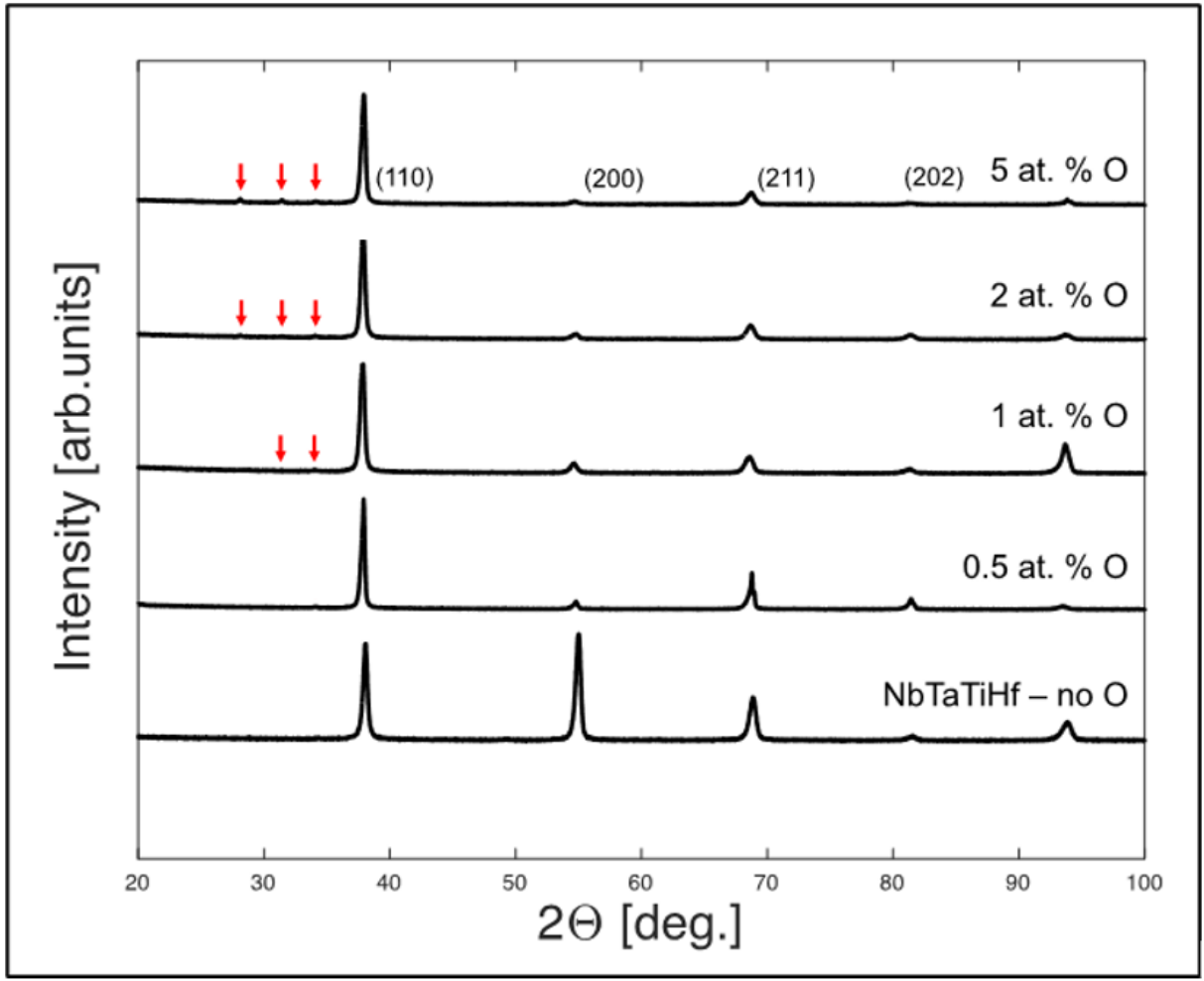

Fig. 2. XRD patterns of the as-cast NbTaTiHf alloys containing targeted contents of O.

X-ray diffraction (XRD) analysis confirmed that all as-cast NbTaTiHf alloys exhibit a dominant body-centered cubic (BCC) phase. In the oxygen-free alloy - which contained a measured baseline oxygen content of ~0.3 at.% - no secondary phases were detected within the resolution of the XRD measurements. However, in alloys with nominal oxygen concentrations exceeding 0.5 at.%, additional diffraction peaks appeared at 2θ angles of approximately 28°, 31°, and 34°. These peaks are consistent with the formation of a secondary oxide phase and align well with reference patterns for monoclinic HfO₂, as reported in prior studies [7–9]. The emergence of these peaks suggests that oxygen concentrations above ~0.5 at.% exceed the solid solubility limit and promote the formation of HfO₂ precipitates within the BCC matrix.

To evaluate the influence of oxygen on room-temperature deformability, the as-cast NbTaTiHf alloys were subjected to cold rolling up to a 90% reduction in thickness or until surface cracking initiated. Representative photographs and corresponding optical micrographs of the rolled specimens are shown in Fig. 3, illustrating the onset and propagation of cracks during deformation. Cracks were observed to propagate perpendicular to the rolling direction, indicating tensile failure along the transverse axis. As the oxygen content increased, the alloys exhibited a marked decrease in cold workability, with all oxygen-containing specimens failing before exceeding a 10% reduction in thickness. This rapid embrittlement is attributed to strong stress localization, either from the formation of secondary oxide phases or from pronounced interstitial strengthening due to dissolved oxygen.

Notably, the alloy containing 0.5 at.% oxygen exhibited a partially ductile fracture morphology, with dimpled and faceted features visible on the fracture surface, suggesting a transitional behavior between ductile and brittle failure. In contrast, alloys with higher oxygen content displayed more catastrophic cracking, consistent with a transition to brittle fracture mechanisms. Due to premature failure during rolling, subsequent microstructural characterization was limited to the as-cast condition, as further thermomechanical processing could not be performed without inducing extensive damage.

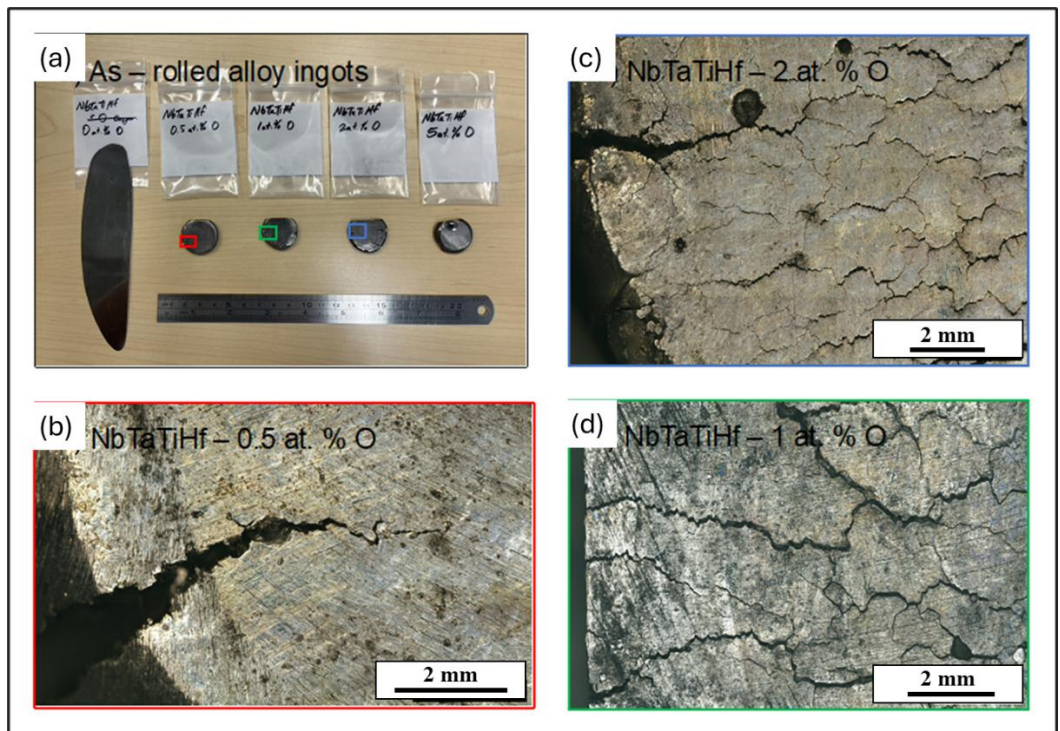

Fig. 3. A (a) photograph of the as-rolled NbTaTiHf alloys, containing 0, 0.5, 1, 2, 5 at. % O added to the alloys, as well as optical micrographs of the as–rolled NbTaTiHf alloys containing (b) 0.5 at. % O, (c) 1 at. % O, and (d) 2 at. % O. The photographs and optical

As shown in Fig. 4, there is no apparent chemical segregation in the NbTaTiHfO0.5 specimen. The line profile of elemental analysis in Fig. 4(c) shows no sign of an increment in O and Hf contents at the grain boundary, suggesting that the O content does not surpass its solubility limit in this alloy. This indicates that mechanical failure is caused by strong random interstitial strengthening and stress concentration localized around interstitial sites. To ensure no nano-precipitates are escaping SEM's observation, additional high magnification S/TEM images (Fig. 5) demonstrate no sign of extra diffraction spots in addition to the typical disordered BCC Bragg spots (as shown in selected area electron diffraction and the fast-Fourier transformation patterns Fig. 5(a)(b). Line profile and elemental mapping in Fig. 5(c)(d) suggest no chemical segregation along the grain boundary, even at high magnification.

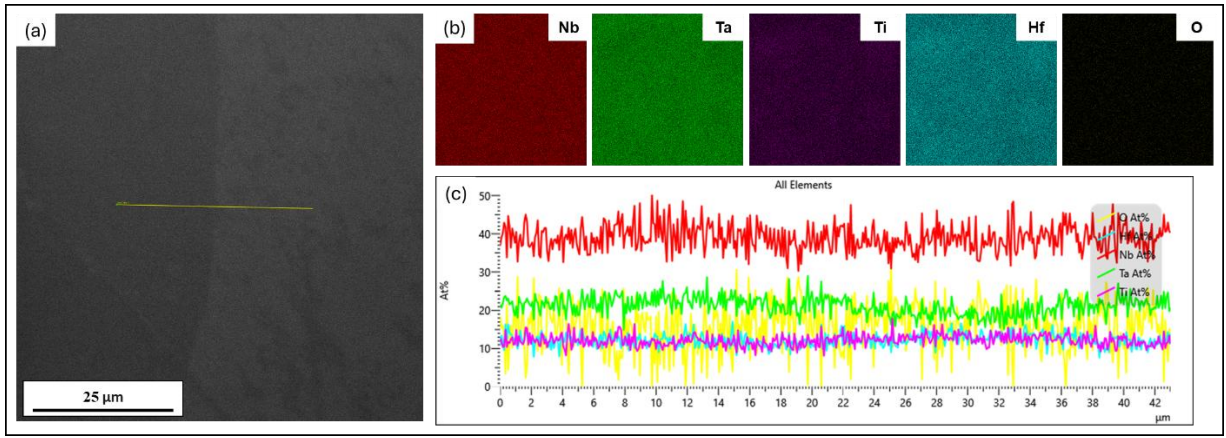

Fig. 4. (a) SEM-SE micrographs of NbTaTiHfO$_{0.5}$ (b) EDS elemental maps. (c) 1-D compositional line profile from the highlighted yellow line in (a).

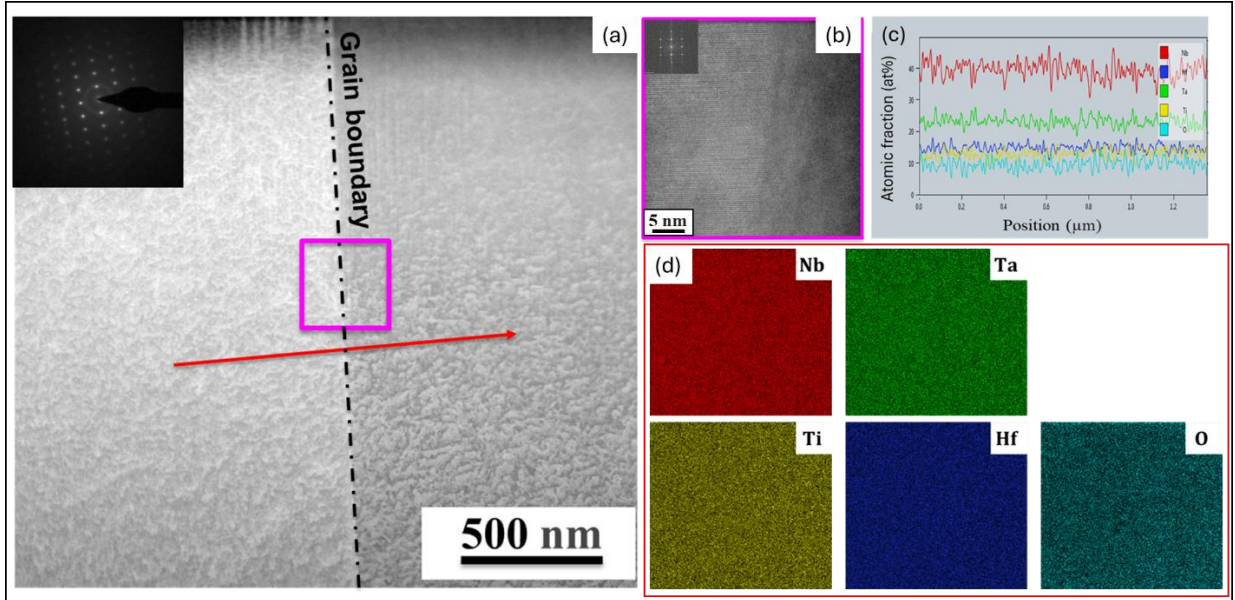

Fig. 5. (a) STEM-BF imaging of NbTaTiHfO$_{0.5}$ and the correlated SAED pattern at the zone axis of [110]. (b) Atomic resolution and the corresponding FFT pattern show no sign of nano-sized precipitates. (c) 1-D compositional line profile from the highlighted red line in (a). (d) EDS compositional map.

At an oxygen content of 1 at.%, pronounced oxide formation is observed along the grain boundaries, as illustrated in Fig. 6, due to the O concentration exceeding the solubility limit of the system which is estimated to be 0.5 at% (see Section 4.2). Additionally, this secondary phase was enriched in Hf and O, indicating the presence of HfO$_2$. These observations are consistent with the XRD, and simulation results revealed in Section 4.2. Further characterization by the EBSD results, as shown in Fig. 13(c-e), demonstrates the formation and the lattice structure of the monoclinic HfO$_2$ ($a$=5.117 Å, $b$=5.175 Å, $c$=5.291 Å, $\alpha$=90°, $\beta$=99.2°, $\gamma$=90°). That is to say, the solubility limit of oxygen in Nb$_{45}$Ta$_{25}$Ti$_{15}$Hf$_{15}$ is established to be between 0.5% and 1%. In response to the rising O content, the morphology of the HfO$_2$ appeared to be different, as shown in Fig. 7 and Fig. 8.

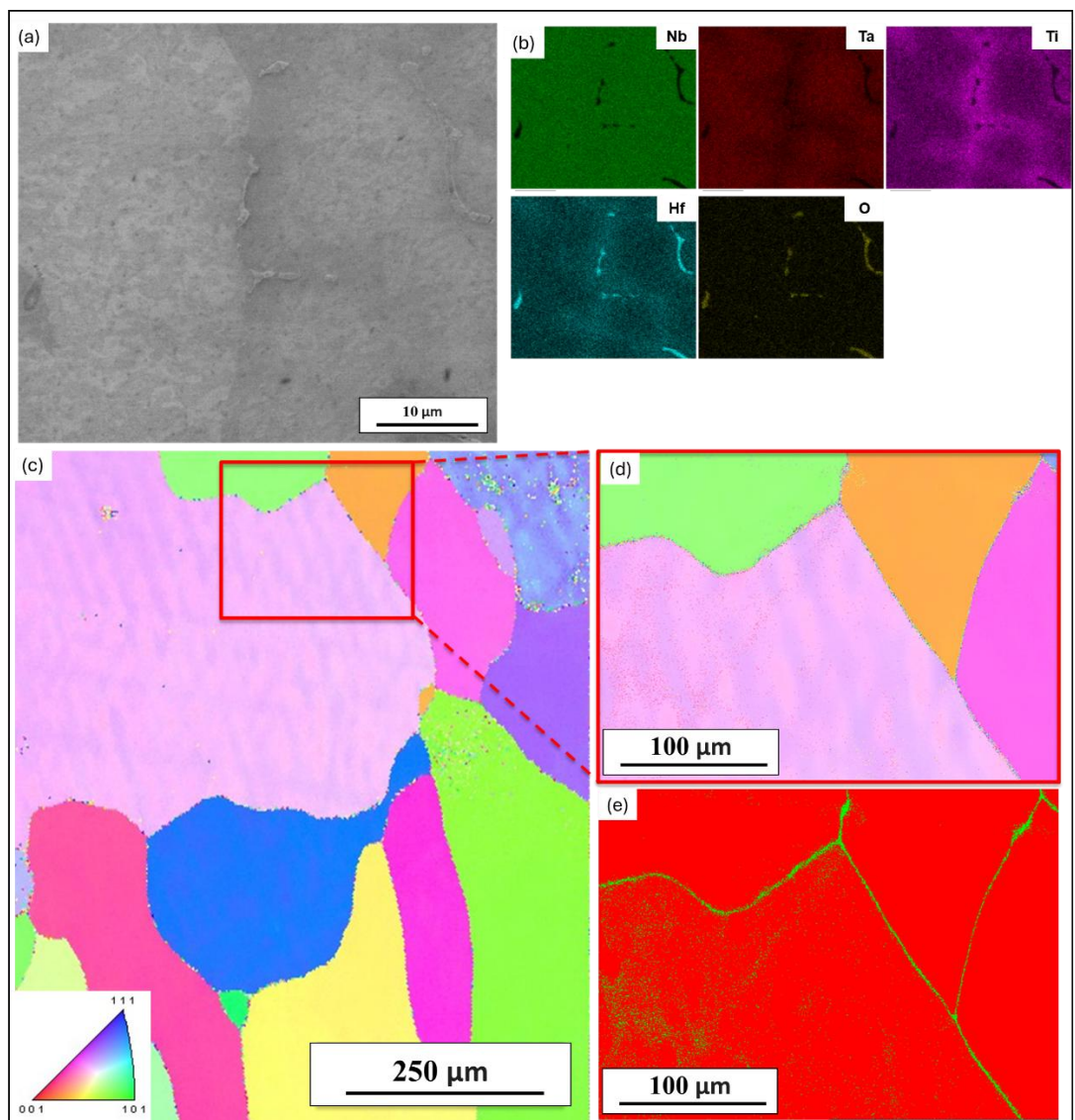

Fig. 6. (a) SEM-SE micrographs of NbTaTiHfO$_1$ (b) EDS elemental maps. (c) EBSD IPF mapping. (d) Enlarged IPF mapping highlighting grain boundary junctions and the corresponding phase map. (BCC and monoclinic HfO$_2$ phases are marked in red and green color, respectively).

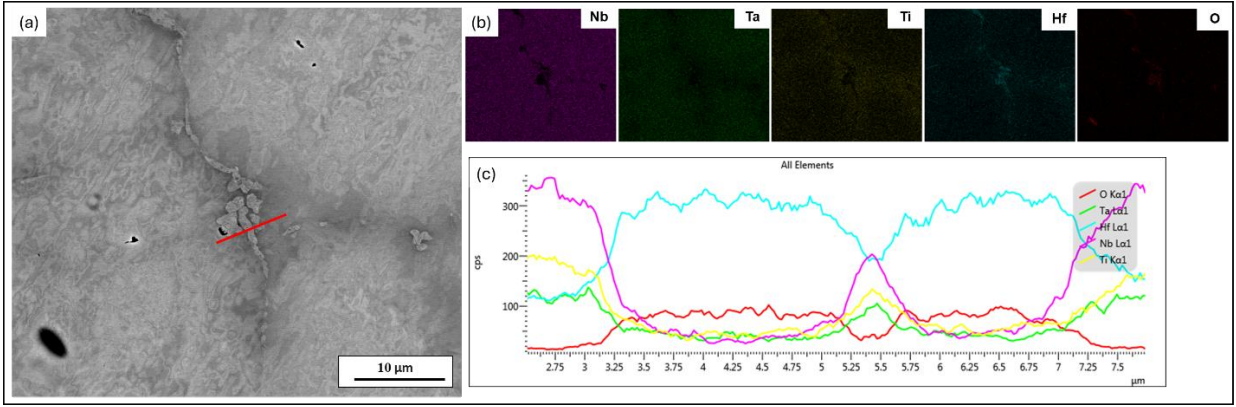

Fig. 7. (a) SEM-SE micrographs of NbTaTiHfO$_2$ (b) EDS elemental maps. (c) 1-D compositional line intensity profile from the highlighted yellow line in (a).

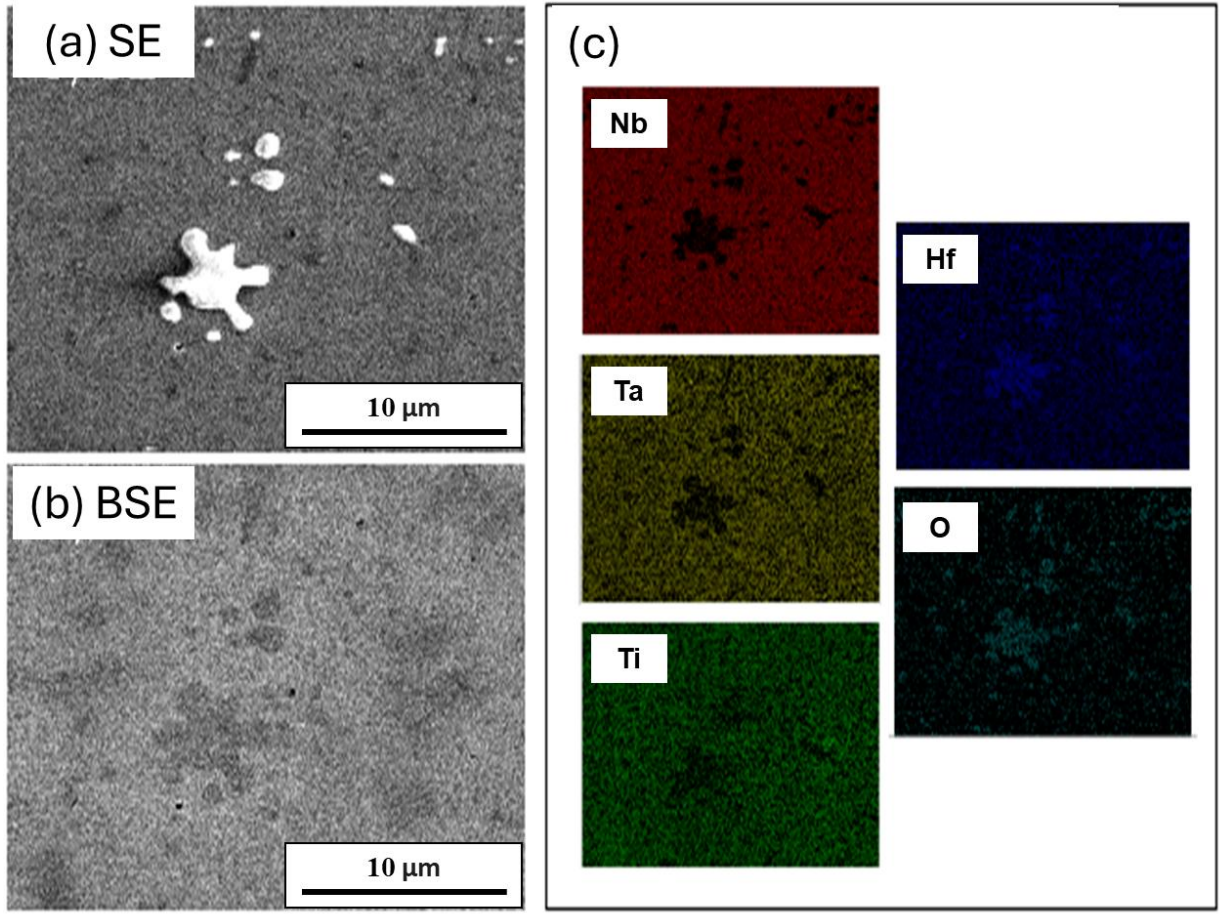

Fig. 8. (a) SE and corresponding (b) BSE micrographs of the NbTaTiHf – 5 at. % O alloy at low magnification, revealing splat-like oxide particles and secondary phases in the BCC matrix of the alloys. (c) EDS elemental maps, further analyzing the oxide particles and secondary phases.

**4.2 Computational Investigation**

In this work, we apply the ORB-v2 model in the *MaterialsFramework* for the structural relaxations and energy calculations. To verify the accuracy of ORB, we first generate 165 BCC structures of Nb-Ti-Ta-Hf with 8 atoms per cell and varying chemical compositions. Additionally, we add an interstitial O atom to each structure to generate another 165 RCCA+O structures. For all the 330 structures, we perform structural relaxations and energy calculations with VASP and ORB. In Fig. 9, we show the free energy differences calculated with VASP and ORB. For most of the structures, the energy differences by ORB and VASP are smaller than 0.1 eV/atom, which is relatively small compared with the energy differences among atomic arrangements in Fig. 10(a). With the verification of ORB results, we can apply it to calculations on larger supercells containing hundreds of atoms, which would be computationally expensive using ab-initio methods.

To study the preference of interstitial O atoms in the $Nb_{0.45}Ta_{0.25}Ti_{0.15}Hf_{0.15}$ RCCA, we generate 500 atomic structures, with a composition of $Nb_{58}Ta_{32}Ti_{19}Hf_{19}$ (in total of 128 atomic atoms per supercell) and 1 O atom. Structural relaxations and energy calculations with and without the O interstitial are carried out using ORB. In Fig. 10(a), we illustrate the total energy of the RCCA+O $E_{RCCA+O}$ versus the total energy of the pure RCCA $E_{RCCA}$ and the energy of O $E_O = E_{RCCA+O} - E_{RCCA}$. The correlation between $E_{RCCA}$ and $E_{RCCA+O}$ is 0.470, while the correlation between $E_O$ and $E_{RCCA+O}$ is 0.852, which shows that the energy of the RCCA+O structure mainly depends on the local chemical environment of the O interstitial instead of the correlation among metallic atoms. For each O atom in the octahedral interstitial space, there are six nearest neighboring metallic atoms. We sort the 500 structures by the number of neighboring metallic atoms of each type of O interstitial and plot this versus the average in Fig. 11. Compared with interstitial O atoms with no Nb or Ta neighbors, the energy is approximately 0.4 eV higher for those without Ti or Hf neighbors. The average energy of O decreases with more Ti and Hf neighbors and increases when surrounded by an increasing quantity of Nb or Ta neighbors. In Fig. 10(b), we show the RCCA+O structure with the lowest free energy per cell among all the 500 structures. The O atom has 3 Hf neighbors, 2 Ti neighbors, and 1 Nb neighbor. Such results indicate the tendency of interstitial oxygen atoms to attract Hf atoms and remain in the interstitial sites surrounded by Hf in the BCC structure, which facilitates the subsequent formation of $HfO_2$.

To better understand the correlations among multi-interstitial O atoms, we develop a Monte Carlo code that optimizes the atomic structure with constrained composition to the one with the lowest free energy. We start the MC process with random bcc structures of RCCA $Nb_{113}Ta_{63}Ti_{37}Hf_{37}$ (250 metallic atoms per cell), with additional 0, 2, 4, and 6 interstitial O atoms. Fig. 12 shows the free energy of the structure as a function of the number of steps. We only show results for 0 and 2 O atoms here to better illustrate the trends. After 200 steps, the free energy converges to within 0.01, and we adopt this as the ground-state energy for that composition. In Fig. 13 we show the atomic structures of RCCA250, RCCA250+2O, RCCA250+4O, and RCCA250+6O, respectively. When multiple O atoms are present, they tend to cluster, and the surrounding region becomes enriched in Ti and Hf, which in turn facilitates the formation of a secondary phase.

To study the formation of metallic oxides in the RCCA+O system, we first consider the stability of each metallic oxide. The oxide formation enthalpies, $\Delta H_f°$, of the constituent elements are shown in Table 2 below, adapted from [33]. With a formation enthalpy of -1108 kJ/mol, $HfO_2$ is the most stable oxide that may form in the alloy, which agrees with experimental observations.

**Table. 2.** Formation enthalpies per mole $O_2$ of the stable oxides of the constituent elements of the NbTaTiHf alloy, showing that the $HfO_2$ is the most stable oxide [33].

| Stable Oxide | $\Delta H_f°$ [kJ/mol] |
|---|---|
| $Nb_2O_5$ | -757 |
| $Ta_2O_5$ | -820 |
| $HfO_2$ | -1108 |
| $TiO_2$ | -941 |

We consider the constraints on chemical potentials for the formation of $HfO_2$,

$$2\mu_O + \mu_{Hf} \geq E_{HfO_2} \qquad (2)$$

As no pure bcc Hf forms,

$$\mu_{Hf} \leq E_{Bcc-Hf} \quad (3)$$

Therefore, a lower limit of chemical potential of O for HfO$_2$ formation is given:

$$\mu_O \geq \frac{1}{2}(E_{HfO_2} - E_{Bcc-Hf}) \quad (4)$$

For HfO$_2$, the $E_{HfO_2}$ is -30.53 eV, while $E_{Bcc-Hf}$ is -9.88 eV calculated with ORB. Therefore, the constraint of HfO$_2$ formation is $\mu_O \geq$ -10.32 eV. It is worth noting that this is only a lower limit estimate of the critical chemical potential that forms HfO$_2$; in other words, it is necessary but not sufficient.

For the chemical potential of O in RCCA, it is a monotonic increasing function of the concentration $\mu_O(x)$. From the structure of RCCA250 to the RCCA250+2O, we have

$$E_{RCCA250+2O} - E_{RCCA250} = \int_0^2 \mu_O(\frac{n}{250})dn \quad (5)$$

With the monotonicity of $\mu_O(x)$, we have

$$\mu_O(0) < \frac{1}{2}(E_{RCCA250+2O} - E_{RCCA250}) < \mu_O(0.8\%) \quad (6)$$

Similarly, we have

$$\mu_O(0.8\%) < \frac{1}{2}(E_{RCCA250+4O} - E_{RCCA250+2O}) < \mu_O(1.6\%) \quad (7)$$

$$\mu_O(1.6\%) < \frac{1}{2}(E_{RCCA250+6O} - E_{RCCA250+4O}) < \mu_O(2.4\%) \quad (8)$$

…

In Table. 3, we give the energy differentials per O atom (such as $\frac{1}{2}(E_{RCCA250+4O} - E_{RCCA250+2O})$ for the line of 4 O atoms). For $\mu_O(0.8\%)$, it is lower than the lower limit of $\mu_O$ forming HfO$_2$. Therefore, we obtain a lower limit estimation of O solubility in Nb-Ti-Ta-Hf alloy with the given composition of 0.8%, which agrees with the experimental results.

.

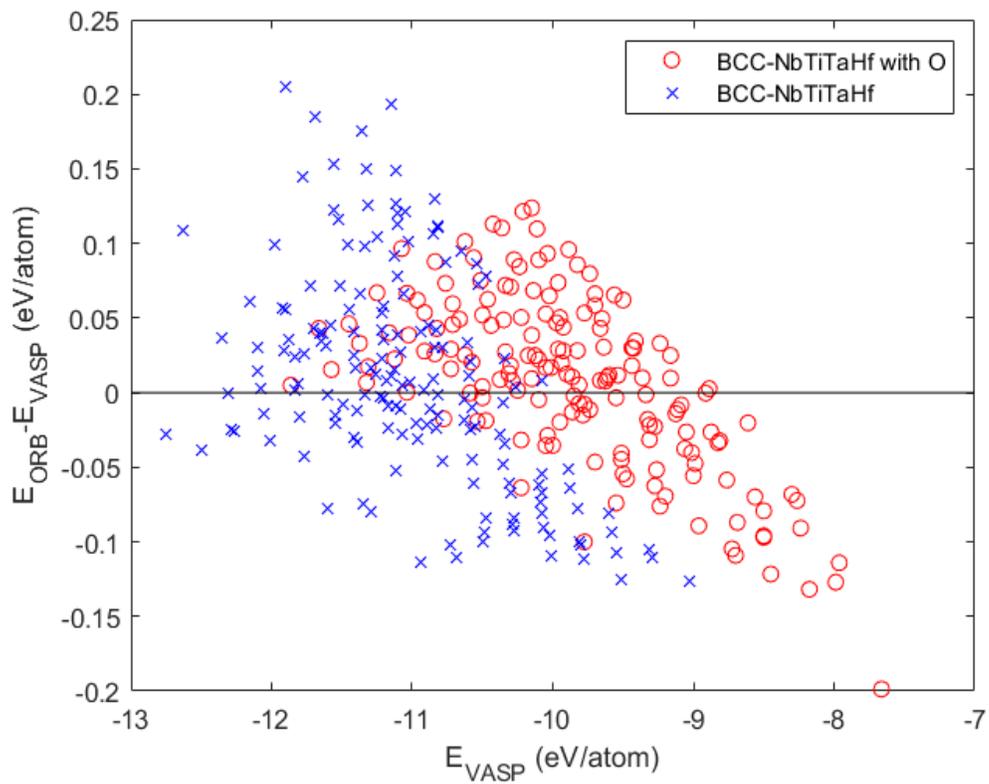

Fig. 9. Energy differences between VASP and ORB calculations for 165 alloy structures with 8 atoms per cell, and 165 structures with 8 metallic atoms and one interstitial O atom per cell.

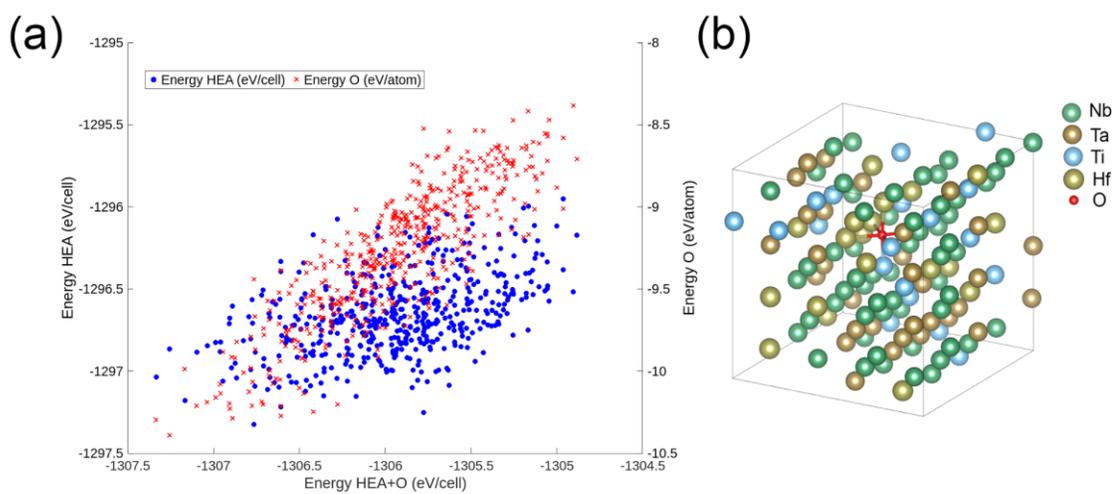

Fig. 10. 128RCCA+O test with 500 random structures. (a) Energy of pure RCCA, energy of RCCA+O, and the energy of interstitial O. The energy of O has a much higher correlation to the energy of RCCA+O; (b)The structure of bcc $Nb_{58}Ta_{32}Ti_{19}Hf_{19}$ + O with the lowest free energy among all 500 structures we calculated. The interstitial O atom has 3 Hf, 2 Ti and 1 Nb neighboring atoms.

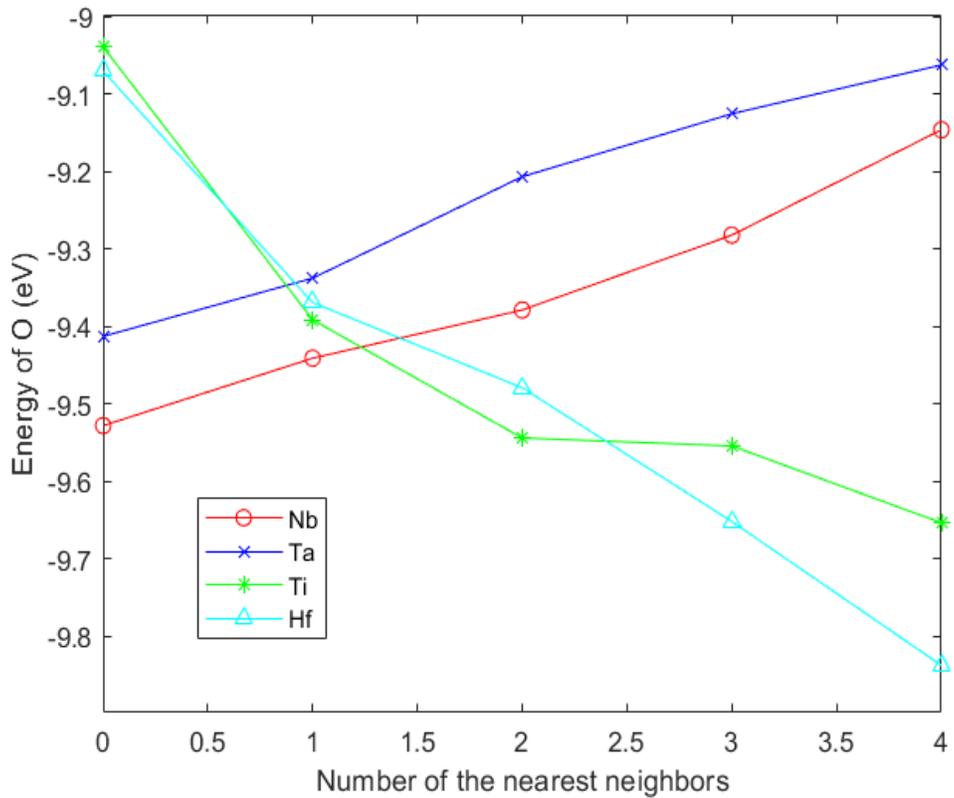

Fig. 11. The average energy of the interstitial O atom versus the number of the nearest metallic atoms. The energy rises with more Nb and Ta neighbors and decreases with more Ti and Hf neighbors.

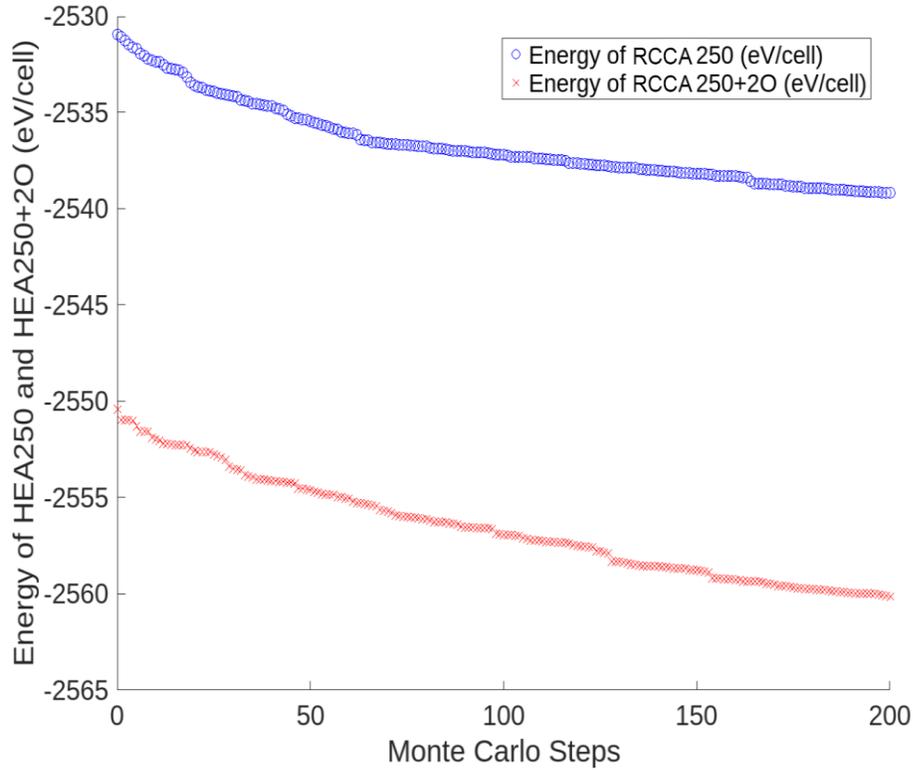

Fig. 12. Monte Carlo process for RCCA250 and RCCA250+2O. The energies converge after 200 MC steps (accepted swaps of either metallic sites or interstitial sites).

Table. 3. Energy of 250RCCA+O structure with O concentration 0 to 2.4% after 200 MC steps.

| Number of O atoms | Concentration of O in RCCA | Energy per cell (eV) | Energy differential (eV per O atom) |
|---|---|---|---|
| 0 | 0 | -2539.31 | - |
| 2 | 0.8% | -2560.51 | -10.60 |
| 4 | 1.6% | -2580.64 | -10.07 |
| 6 | 2.4% | -2598.43 | -8.90 |

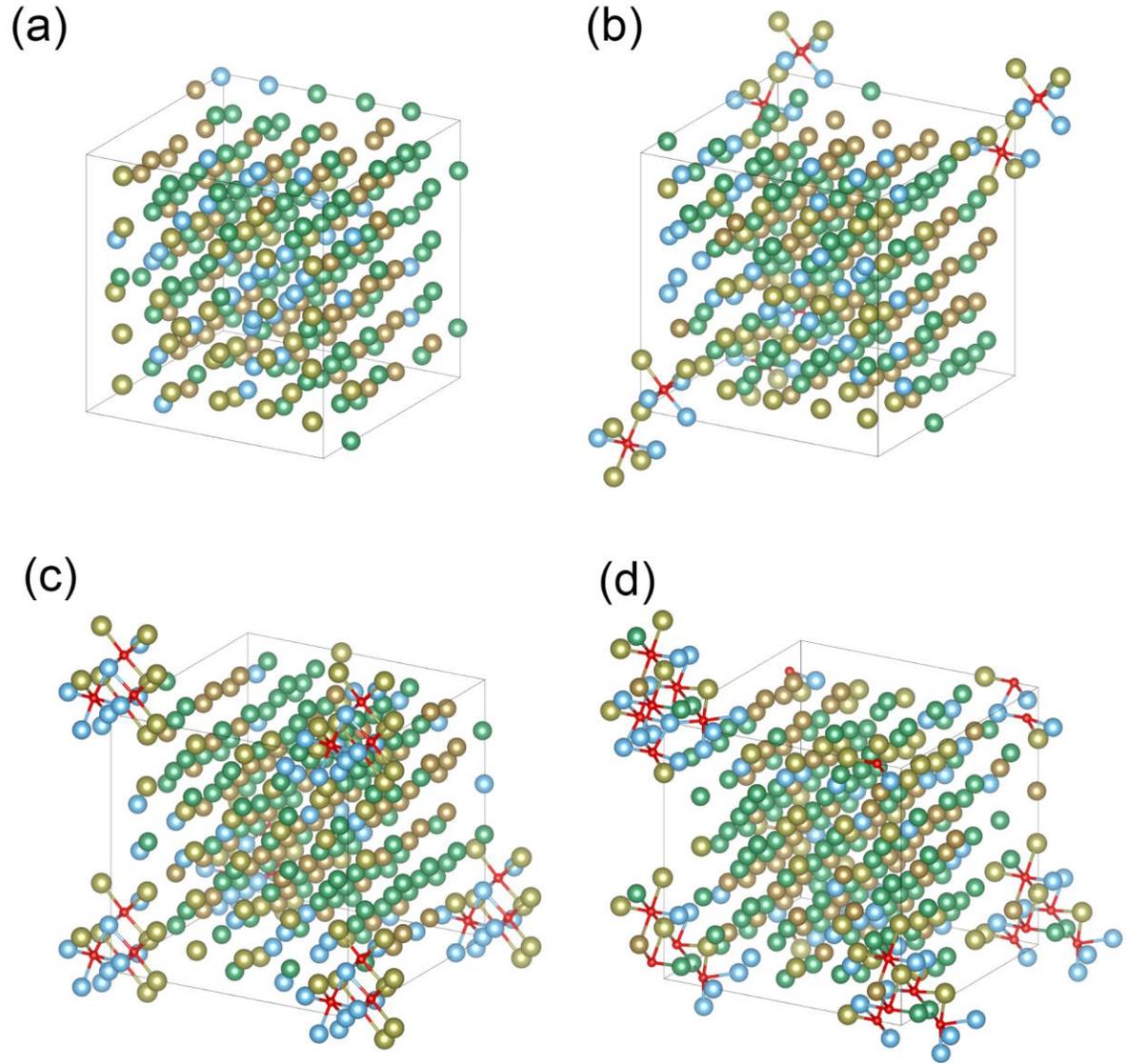

Fig. 13. Atomic structures of RCCA250+O optimized with 200 MC steps. (a) with no interstitial O; (b) with 2 O atoms per 250-atom cell; (c) with 4 O atoms per 250-atom cell; (d) with 6 O atoms per 250-atom cell.

## 5. Conclusion

In this work, we investigated the solubility limit of O in $Nb_{45}Ta_{25}Ti_{15}Hf_{15}$ RCCA, which was determined to be between 0.8 and 1.0 at%. The following conclusions can be drawn from this work :

(a) The oxygen interstitial atoms were found to behave differently, corresponding to their concentration in the $Nb_{45}Ta_{25}Ti_{15}Hf_{15}$ system. Given the characterization results, they are prone to occupying interstitial sites when their concentration is 0.5 at% or lower, whereas a secondary $HfO_2$ phase is promoted when their concentration reaches 1 at% or higher, leading to distinctive fracture behaviors.

(b) Using Monte Carlo simulations with MLIP, we determined the lower limit of oxygen solubility in the $Nb_{45}Ta_{25}Ti_{15}Hf_{15}$ RCCA to be 0.8 at%. Below this concentration, the chemical potential of oxygen in the RCCA interstitial site is lower than that required for $HfO_2$ formation.

On the computational side, a notable advancement of this study is the integration of universal MLIPs and Monte Carlo simulations to model the interstitial oxygen in RCCAs, with high efficiency and accuracy. By benchmarking ORB-v2 against DFT, we confirmed its accuracy and reliability for describing RCCA+O systems. Large-scale sampling of 128-atom supercells revealed the preferred local chemical environments of interstitial oxygen. Furthermore, integrating MLIP with Monte Carlo simulations enabled us to explore structural configurations and estimate the oxygen solubility limit in NbHfTaTi RCCA. These results highlight the efficiency and capability of MLIP-based methods to overcome the limitations of DFT, offering new opportunities to investigate complex alloy–interstitial interactions and possibly expand their scope to other interstitial elements.

In future studies, our methodology can be further extended to investigate other RCCAs and additional interstitial elements beyond oxygen, such as carbon and nitrogen, enabling systematic evaluation of solubility limits and the formation of secondary phases across different alloy systems. Moreover, our computational workflow enables efficient exploration of compositional tuning of RCCAs by carefully adjusting the relative proportions of constituent metallic elements to enhance oxygen solubility. Such compositional strategies may provide a pathway to improve the mechanical properties of RCCAs, offering valuable guidance for the design and optimization of advanced high-temperature structural materials.

**CRediT authorship contribution statement**
**Aomin Huang:** Writing – review & editing, Writing – original draft, Visualization, Validation, Methodology, Investigation, Conceptualization. **Siya Zhu:** Writing – review & editing, Writing – original draft, Visualization, Validation, Methodology, Investigation, Conceptualization. **Calvin Belcher:** Writing – original draft, Investigation. **Ryker Rigsby:** Investigation. **Raymundo Arróyave:** Writing – review & editing, Writing – original draft, Supervision, Software, Resources, Project administration, Methodology, Investigation, Funding acquisition, Formal analysis, Conceptualization. **Enrique J. Lavernia:** Writing – review & editing, Writing – original draft, Supervision, Software, Resources, Project administration, Methodology, Investigation, Funding acquisition, Formal analysis, Conceptualization.

**Declaration of competing interest**

The authors declare that they have no known competing financial interests or personal relationships that could have appeared to influence the work reported in this paper.

**Acknowledgement**

The authors acknowledge the support provided by the State of Texas Governor's University Research Initiative (GURI), the Texas A&M University Chancellors' Research Initiative (CRI), and the National Science Foundation Materials Research Science and Engineering Center program through the UC Irvine Center for Complex and Active Materials. RA and SZ acknowledge the support from the U.S. Department of Energy (DOE) ARPA-E ULTIMATE Program through Project DE-AR0001427 and DE-AR0001860. RA also acknowledges NSF through their DMREF Program under Grant No. 2119103. Calculations were carried out at the Texas A&M High Performance Research Computing (HPRC) Facility.